\documentclass[format=acmsmall,authorversion]{acmart}
\pdfoutput=1
\usepackage{graphicx}
\usepackage{url}
\usepackage{caption}
\usepackage{booktabs}
\usepackage{multirow}
\usepackage{tabularx}

\usepackage{tikz}

\newcolumntype{Y}{>{\centering\arraybackslash}X}

\newcommand*\circled[1]{\tikz[baseline=(char.base)]{\node[shape=circle,draw,inner sep=.75pt, line width=.7pt] (char) { #1};}}
\newcommand*\dcircled[1]{\tikz[baseline=(char.base)]{\node[shape=circle,draw,inner sep=.75pt, dashed, dash pattern = on 1.5pt off .75pt, line width=.7pt] (char) { #1};}}

\newcommand{\changed}[1]{#1}

\setcopyright{none}
\renewcommand\footnotetextcopyrightpermission[1]{}
\acmDOI{}

\begin{document}

\title{Code Renewability for Native Software Protection}

\author{Bert Abrath}
\email{bert.abrath@ugent.be}
\affiliation{%
  \institution{Ghent University}
  \department{Computer Systems Lab}
  \country{Belgium}
}

\author{Bart Coppens}
\email{bart.coppens@ugent.be}
\affiliation{%
  \institution{Ghent University}
  \department{Computer Systems Lab}
  \country{Belgium}
}

\author{Jens Van den Broeck}
\email{jens.vandenbroeck@ugent.be}
\affiliation{%
  \institution{Ghent University}
  \department{Computer Systems Lab}
  \country{Belgium}
}

\author{Brecht Wyseur}
\email{brecht.wyseur@kudelskisecurity.com}
\affiliation{%
  \institution{NAGRA-Kudelski Group}
  \city{Lausanne}
  \country{Switzerland}
}

\author{Alessandro Cabutto}
\email{a.cabutto@uel.ac.uk}
\affiliation{
  \institution{University of East London}
  \department{Department of Computing and Engineering}
  \country{United Kingdom}
}

\author{Paolo Falcarin}
\email{falcarin@uel.ac.uk}
\affiliation{
  \institution{University of East London}
  \department{Department of Computing and Engineering}
  \country{United Kingdom}
}

\author{Bjorn De Sutter}
\email{bjorn.desutter@ugent.be}
\affiliation{%
  \institution{Ghent University}
  \department{Computer Systems Lab}
  \country{Belgium}
}

%
\renewcommand{\shortauthors}{Abrath et al.}

\begin{abstract}
Software protection aims at safeguarding assets embedded in software by
preventing and delaying reverse engineering and tampering attacks. This paper
presents an architecture and supporting tool flow to renew parts of native
applications dynamically. Renewed and diversified code and data belonging to
either the original application or to linked-in protections are delivered from
a secure server to a client on demand. This results in frequent changes to the
software components when they are under attack, thus making attacks harder. By
supporting various forms of diversification and renewability, novel protection
combinations become available, and existing combinations become stronger. The
prototype implementation is evaluated on a number of industrial use cases.
\end{abstract}



\maketitle

\section{Introduction and Motivation}
\label{sec:introduction}



\textit{Man-At-The-End} (MATE) attackers use
debuggers, emulators, custom operating systems, analysis tools, etc.\ to reverse
engineer or tamper with software distributed by providers of software, service, and content.
\changed{The Global Online Piracy study~\cite{GOP} shows the continuous worldwide presence of online piracy of digital contents, such as movies, music, and games, while the latest BSA Global Software Piracy Study~\cite{BSA} states that 37\% of software installed on computers worldwide is not licensed,  amounting to \$46.3 billion in losses due to software piracy.
The same study shows that malware often spreads through unlicensed software distributed on the internet, causing a wider number of security attacks and consequent revenue losses; cyber criminals are now targeting mobile apps as well: malware variants on mobile devices increased by 54 percent last year, with 24,000 malicious mobile apps blocked every day~\cite{ISTR}.} 

Software protection techniques aim at protecting the
integrity and confidentiality of the provider's assets in the software by making
it harder to reverse engineer and tamper with~\cite{collberg2009book,FalcarinCAJ11}.
\changed{In unprotected applications, this is all too easy. In 2016 Arxan, one
of the major vendors of software protection solutions, put forward that 98\% of
mobile apps lack binary code protection and can be easily reverse-engineered
and tampered with~\cite{Arxan}. The use of protections to mitigate this issue
is becoming increasingly popular, however. In 2017 Gartner projected that in
2020, 30\% of enterprises will use software protection to protect at least one
of their mobile, IoT, and JavaScript critical applications~\cite{Gartner}.}

Each individual protection technique only affects a small set of attack vectors, and
applying only a few will merely divert the attacker's attention to the
remaining unprotected attack vectors. Thus, multiple techniques need to
be combined to ensure all these possible paths-of-least-resistance are hardened.
Overall, protections aim for (i) increasing the effort needed to
identify successful attack vectors; (ii) increasing the effort
needed to manually exploit these attack vectors (iii) increasing the effort needed to
automate and scale-up their exploitation; (iv) minimizing the number of 
instances on which automated attacks can be deployed; (v) reducing the window
of opportunity for generating income from an attack.

Protections hence need to be diversified, such that they maintain a level of
resilience and different versions can be generated of the same
functionality. Defenders need a mechanism to renew (i.e., update) assets and
protections in the field such that the attack vector identification has to be
re-done frequently, the value of assets decreases rapidly, and the protections'
behavior varies over time. If temporal variation is unpredictable, attackers
always need to take into account all protections to remain undetected and
successful. This furthermore means that not all protections need to be
active at the same time, which can allow the run-time overhead to remain acceptable.

This paper presents the ASPIRE renewability framework for delivering
renewability to the native executables and libraries that often implement the
security-sensitive functionality of always-connected mobile applications. \changed{Other types of
  native software can be targeted as well, as long as the deployed protected versions can be assumed to be
  always connected. The renewability framework is part of the broader ASPIRE framework that consists of a protected application
  architecture and the compiler tool flow that supports the automated deployment of
  protections fitting that architecture.} The renewability part of that overall framework leverages existing diversity
techniques~\cite{Larsen14} and protection techniques to generate the variation required for renewability
variation. This paper's main contributions are:
\changed{
\begin{enumerate}
\item The protected application architecture that supports
  many forms of software renewability and the composition of those forms with other protections. It builds on existing ideas from literature~\cite{falcarin2011exploiting,collberg2012distributed}, but the proposed design is more mature and is the only academic effort that has actually been validated by industrial security architects on case studies of real-world complexity~\cite{D106}.
\item The compiler tool flow that supports the automated deployment of many protections, i.e., that injects renewable and other protections into applications to instantiate the protected application architecture automatically. Whereas the designs pitched and evaluated in literature only focused on one form of protections~\cite{falcarin2011exploiting,collberg2012distributed}, our tool flow is more feature complete. Most importantly, it supports many different protections and compositions thereof as discussed later in the paper.  That capability has been validated in the industrial effort mentioned above. 
\item A discussion of a number of applications of the renewability framework, i.e., concrete forms to mitigate specific attacks by making existing protections stronger through renewability. The discussed forms are not comprehensive or exhaustive, but they illustrate the potential of the renewability framework to strengthen existing protections. 
\item The evaluation of a prototype implementation of the tool flow and the discussed forms of renewability. This prototype was also part of the mentioned industrial validation effort. Large parts of it are available as open source for future research and reproducibility. No other, such feature-complete tool flow has been presented in literature or is available as open source.
\end{enumerate}
}

Section~\ref{sec:attack_model} discusses the MATE attack model.
Section~\ref{sec:architecture} presents the overall framework design and
architecture, after which Sections~\ref{sec:renew_polic} and~\ref{sec:mobile_data} discuss specific
features. Section~\ref{sec:toolflow} presents the tool flow to support
automated deployment of the framework. Section~\ref{sec:applications} discusses
concrete uses of the framework to mitigate a variety of concrete MATE attack
steps. Section~\ref{sec:evaluation} evaluates the proposed renewability
framework and the prototype implementation in terms of robustness, overhead, and
scalability. After related work is discussed in Section~\ref{sec:related},
conclusions are drawn in Section~\ref{sec:conclusions}.

\section{Attack Model}
\label{sec:attack_model}

We aim to protect software against MATE attacks. In their labs, MATE attackers
have full access to---and full control over---the software under attack, as
well as over the system on which the software runs. They can use static
analysis tools, emulators, debuggers, custom operating systems and all kinds of
other hacking tools. The attacks are looking to break the integrity and
confidentiality requirements of assets, e.g., to steal keys or intellectual
property, and to break license checks. They do so by means of reverse
engineering and by tampering with the code and its execution.

We focus on mobile applications distributed by providers of content, software,
and services. Often, their GUI parts are implemented in managed languages such as
Java. Because of the ease with which, e.g., Java bytecode can be
reverse-engineered, and because of performance concerns, the security-sensitive
assets are typically still implemented in dynamically linked, native libraries
that are packaged with, e.g., the Java apps. The software under attack
therefore consists of native binary files (this includes both dynamically linked 
libraries as well as stand-alone executables). Because of the economic value of the
assets, we assume software protection techniques are deployed in and on the native
code~\cite{collberg2009book}.

We only target always-online applications, such as video streaming apps or edge apps
that connect to cloud servers. While this is a limitation, the omni-presence of
wireless networks (4G, 5G, WiFi) has resulted in a big enough market to develop
protections that exploit the always-online feature.

Our protections target economically driven attackers. We aim at increasing
their attack investment cost, at lowering their profit, and at tilting the
balance between the two. The protection is effective when the attackers expect
a negative return on investment before they even start the attack or while they
are still pursuing it, as well as when they expect a higher return on
investment from attacking other providers' software. The protections then
stopped the attackers before they had a chance to succeed. Even if the
attackers succeeded, though, the protections can have delayed them enough for
the provider to make a healthy profit of the assets. In that case as well, the
protections can be considered successful.

In their lab, MATE attackers execute an attack strategy and a series of attack
steps. The strategy is adapted on the fly, based on: the results of previous
attack steps; hypotheses that the attackers formulate and test regarding
assets, deployed protections, other relevant features of the software under
attack (such as the locations of relevant code and data); and the perceived
path of least resistance. With the perceived path of least resistance, we mean
the sequence of future attack steps that the MATE attackers consider the most
efficient and effective to pursue given their expertise, skills, and tool
availability. We refer to existing literature for more information and models
of the attack processes as obtained through empirical experiments with various
kinds of attackers on various kinds of assets~\cite{emse2019}. In the context
of this paper, one important aspect to point out is that in the eyes of MATE
attackers, many seemingly uninteresting artifacts of software (system calls,
control flow structures, ...) are in fact interesting, because they can serve
as hooks for the attackers to guide their search to the really interesting
code.

To be effective, protections should cover as many attack paths as possible that
might be paths-of-least-resistance for certain attackers. The protections can achieve this
by making the individual attack steps on the paths more expensive or
time-consuming, by requiring extra attack steps, or by preventing certain
attack steps and the automation thereof. Section~\ref{sec:applications} will
discuss several concrete attack steps against which protections exist that can
be made more effective by making them renewable with the presented framework
and architecture. In general, these steps are attack vector identification and
attack vector exploitation steps that require a certain amount of
repeatability, such as the iterative development and later use of customized
scripts that work well as long as the software they operate on remains the
same.

It is commonly accepted that sufficient protection can only be achieved by
combining many protections in a layered fashion. The deployed protections then
become assets themselves, protecting the original assets, the artifacts that
attackers can hook onto, and each other. The value of the proposed renewability
framework and architecture hence cannot be judged in isolation. The supported
forms of renewability are supposed to be combined with other protections that
protect against additional attack vectors, and that protect the components of
the renewability implementation. The communication to a secure server to
download renewed assets and protections, for instance, is supposed to be protected
by sufficiently strong cryptography, of which the keys are protected through
white-box cryptography, of which the code is obfuscated to prevent static
reverse engineering, and anti-debugging techniques to protect against dynamic
reverse engineering. Similarly, remote attestation is supposed to be used for
hampering replay attacks, i.e., for checking that a client application actually
executes freshly downloaded code rather than old copies stored on disk by an
attacker.

In the ASPIRE project, we reached the necessary composability of renewability
with other protections in an open-sourced protection tool
chain~\cite{aspire_toolflow}. The renewability framework and architecture
presented here are only one of several novel aspects of that tool
chain. Composability of all kinds of protections in the tool chain is out of
this paper's scope.

Our MATE attack model neglects hardware-based protections. Off-the-shelf
processors offer limited protection against MATE attacks. SGX enclaves can leak
information in contexts similar to MATE
attacks~\cite{sgxcacheattacks,branchshadowingsgx}. Furthermore, they are
restricted in their interaction with outside components, so they cannot protect
all code. TrustZone~\cite{trustzone} is only effective in well-configured
systems. In a lab, a MATE attacker can easily disable the protection.
Furthermore, many lower-end devices lack hardware protection. For those,
software-only protection is the only available option. Moreover, hardware-based
protection is considered a risk by some, because it is expensive and at the
same time not renewable~\cite{counterpoint}. The reason for this is that when a 
hardware defense mechanism is broken at some point, e.g., because implementation
bugs are discovered, it is typically very hard---if not impossible---to fix it,
so all systems relying on that hardware are vulnerable from then on. Software
renewability offers a complementary solution for such scenarios.

\section{The ASPIRE Renewability Architecture}
\label{sec:architecture}

\begin{figure*}
\centering
\includegraphics[width=\linewidth]{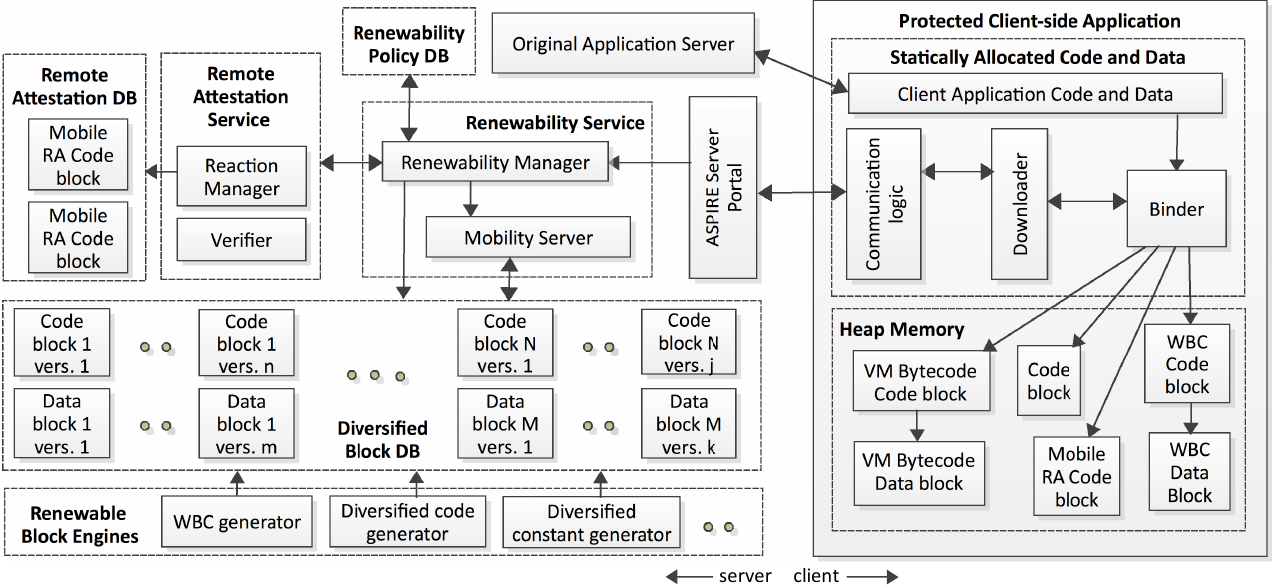}
\Description{ASPIRE Renewability Architecture. On the right, the native application
  from which static code and data has been extracted and to which code mobility
  components have been added. On the left, the original application server in
  case the application was an online application, and all server-side components
  that implement the many forms of renewability.}
\caption{ASPIRE Renewability Architecture. On the right, the native application
  from which static code and data has been extracted and to which code mobility
  components have been added. On the left, the original application server in
  case the application was an online application, and all server-side components
  that implement the many forms of renewability.}
\label{fig:arch}
\end{figure*}

Figure~\ref{fig:arch} visualizes the ASPIRE renewability architecture. It is based
on code and data mobility, which builds on the existing concept of code
mobility~\cite{cm}.
From the binary file of a client app or library that needs to be protected,
parts of the statically allocated code and data sections are extracted. These
parts correspond to code that will need to be renewed dynamically, i.e., when
the app or library is actually running. By simply removing this code and data,
it is already protected against purely static MATE attacks. The code and static
data (Client Application Code and Data in Figure~\ref{fig:arch}) that remains in
the binary is extended with support code: Communication Logic, a
Downloader, and a Binder.

The Communication Logic implements protection-agnostic communication 
and protocols to the Server Portal. Its prototype implementation offers a simple
request protocol and a WebSockets protocol~\cite{ws} for protections that
need occasional connections and/or server-initiated communications and a persistent connection.
(Replacing WebSockets with a more
secure implementation is orthogonal to this paper.)

The Downloader implements the communication. Upon requests from the Binder, it connects to the Mobility Server to download mobile code and data blocks. The downloaded
blocks are then mapped at randomized locations on the heap of the
running client. In basic code mobility, these blocks
correspond to individual code fragments extracted from the statically
allocated code of the protected app. 


The Binder initiates the download requests on demand and ensures that all
control transfers and accesses to and from downloaded code and data execute
correctly. Each transfer \emph{into} a mobile block is redirected via a stub
that transfers control to the block's address, which it finds in a Global Mobile
Redirection Table (GMRT). Until a block has been downloaded, the found
address is that of another stub that invokes the Downloader with the correct
input and that performs the necessary allocation and bookkeeping. This includes
replacing the stub's address with the block's address in the GMRT, and then
continuing execution at the entry of the block. All transfers \emph{out of} a block are
transformed into \emph{offset-independent code} by adding a level of
indirection~\cite{cm}. 

In basic code mobility, a downloaded mobile block remains mapped on the heap of the running
process until it halts. To support advanced forms of renewability, we
extended the Binder to support the flushing of mapped blocks and subsequent
re-downloading of renewed, different versions of those blocks. We also extended it
to support mobile data, which is necessary to support several
interesting forms of renewability that will be discussed later. 

The protection-agnostic ASPIRE Server Portal forwards communications between
clients protected with (multiple) online protections and the corresponding
services. In our prototype, it also supports
client-server code splitting~\cite{bs} and remote attestation (RA) 
techniques~\cite{Viticchie16}. 

The Renewability Manager selects which mobile code and data blocks need to be
delivered to a running client. By varying the mobile block versions that are
delivered to different clients and at different times, the assets and
protections implemented by that mobile code and data can be renewed. The Mobility Server takes care of the actual delivery and interaction with the
Downloader in stateless communication: The
Server does not keep track of existing sessions with clients for the
sake of scalability; it just serves the right block whenever a request arrives
from one of the clients, based on the policy implemented by the Manager.

The mobile code and data blocks are stored in a database (DB). That Diversified
Block DB can hold multiple, diversified versions of each block. For most
forms of renewability, the different mobile blocks and the different versions
thereof, are independent of each other. This is the case because either all
versions of a block implement exactly the same semantics, or because one block's
semantics is independent of the other blocks' semantics. For some forms of
renewability, however, there may exist dependencies between the blocks. Some interesting cases are discussed later. 

Different server-side code generators (Renewable Block Engines) produce
diversified mobile blocks.  Depending on the renewability policies, these
generators can generate blocks a priori or on demand. For example, in case a
policy only aims at delivering different versions of a block with the exact same
semantics, the DB can be populated a priori. If specific versions need to be
generated to react to events, they can instead be generated on
demand. Obviously, if the event to react to is an actual request from a running
app, the on-demand generation will result in a higher response time.

The renewable code engines are application-dependent, as they generate mobile
blocks matching the code and data fragments that were extracted from the static
binary of the client software. Section~\ref{sec:toolflow} discusses how
these engines are generated.
Obviously, but not drawn in Figure~\ref{fig:arch} to keep it simple, the
Renewability Manager also has to interact with the code engines to know what
code is in the DB, and to trigger on-demand generation of blocks.

Furthermore, the Renewability Manager can interact with other protection
servers. Figure~\ref{fig:arch} includes an example Remote Attestation
Service. That interaction
can be exploited in two ways.
First, the renewability policy itself can interact with the other protection
server. In the case of RA, the interactions can involve notifications of failed
attestations, and communication about the mobile blocks that were delivered to
the client such that the RA can attest them.
Secondly, the other online protection might also have client-side protection
components, such as specific hash functions used in RA code guards that need to be delivered as mobile blocks via the Renewability Manager. Figure~\ref{fig:arch} shows this for mobile RA blocks. Note that the difference between the Mobile RA Code
blocks and the Code and Data blocks in the Diversified Block DB is that the
former are application-independent components of a deployed protection, while the
latter implement original client-side functionality.





\section{Integrating Renewability into Existing Applications}
\label{sec:renew_polic}

In order to integrate renewability into existing applications, there are two 
choices that need to be made: (1) where and how decisions will be made to renew
blocks, and (2) decide how these decisions will be communicated with the client
application. We call the former the renewability policies, and the latter the
renewability communication design. We will now discuss the spectrum of options
and trade-offs that can be made for both kinds of policy.

\subsection{Renewability Policies}

Renewability policies define when a client needs to discard and replace
downloaded mobile blocks with renewed ones. The decision to renew a block can
either be made server-side, or it can be made in the client itself. When
considering client-side decisions, these can be made either without external
inputs (the logic is set in stone), or it can be that the application
implements a policy that has been dictated to it by the Renewability Server.
Either way, a MATE attacker might be able to learn or even subvert the
renewability policies. A major advantage of server-side decisions, is that the
client then only learns about the concrete decisions taken by a policy (i.e.,
the flushing commands), and not about the rules that lead to these decisions.
On top of that, a persistent, server-initiated connection to pass policy
decisions to the client enables dynamic policies that can be adapted on the
fly. A compromise between these two approaches would be to send a policy
description to the client with each served block. A policy would then be
immutable in between the delivery of blocks. In the rest of this section we
will consider server-based renewability policies.

When the renewability policy is implemented on the server, we can either make
this an application-agnostic \emph{decoupled policy}, or a \emph{coupled
policy} that is tightly integrated with the Original Application Server.

In the case of decoupled policies, no changes need to be made to the
application source code when compared to the original, non-renewable ASPIRE
architecture: It suffices to add annotations in the source code. The decision
of whether and when to renew certain blocks is made completely independently
from the application, by the server-side Renewability Manager. When this
component decides to renew a certain block in a certain application instance,
it sends a flushing command to that application instance.

Implementing policies in a decoupled manner on the renewability server somewhat
restricts the manner in which the renewability policies can react to events
occurring in the protected application. We would however argue that there is
still quite a lot of leeway left to react: We can compose different
(application-agnostic) protection techniques and change the policies based on
the state and observations of these other protection techniques. For example,
in our prototype implementation, the integrity violations that are observed by
the remote attestation component are passed on to the renewability server,
which changes its policy based on these observations. Several reactions are
possible: the Mobility Server can stop serving blocks, the client can be
notified in the next communication through the ASPIRE Server Portal, or the
Original Application Server can be informed that it should stop delivering
content~\cite{Viticchie16}. Furthermore, it has already been demonstrated in
the ASPIRE project that other protections, such as client-server code
splitting, can be used to let a protection server keep track of different
events in the client~\cite{Viticchie16,bs}. Decoupled policies thus lead to a
clear separation of concerns.

Alternatively, in a \emph{coupled} policy, the (server-side) application logic
is tightly integrated with the renewability policy. The decisions of which
(specific instances of) blocks to flush can be based explicitly on the state in
which a specific client happens to be, and can be made to coincide with other
actions that are taken in the application server. For example, in the case of a
streaming video application, the streaming server can be integrated in the
renewability policy so as to force the client to download a different decoder
function after a specified number of video frames have been sent. The
application server can thereafter send differently encrypted or encoded frames,
which the old decoder function is not able to decode. This option offers the
vendor much more control over the renewability policy. The price, however, is a
sharp dent in the separation of concerns, as the protection is now to a large
degree hard-coded in the application source code.

\subsection{Renewability Communication Design}

After a decision has been made by the renewability policy, it has to be
communicated to the protected application. We elaborate on two possible designs
for this communication: an application-agnostic, decoupled communication
design, and a tightly-integrated, coupled communication design.

In the case of a decoupled communication design, we can build on existing the
ASPIRE components: The client-side Binder component handles flushing commands
received from the server, while the server-side Renewability Manager sets up a
bi-directional connection with the application for future, server-initiated
flush requests. Flushing consists of the deallocation of individually
specified---or even all---mobile blocks, the resetting of addresses in the
GMRT, and informing the server of its completion. In this manner, the server
can be aware that flushing is not happening, and suspect the client is being
tampered with. When the client fails to confirm the flush request within a
given time, an appropriate reaction can be activated.

Conversely, in a coupled communication design, both the protected application
logic and server logic (including the existing protocols) are augmented in
order to support all communication logic that is required for renewability
(e.g., receiving new blocks, receiving and confirming flush requests, etc.).
The application server needs to communicate directly with the Renewability
Service to obtain mobile blocks, and will need to embed those blocks---together
with descriptions of renewability actions---in the packets sent to the client
application. This is practical, e.g., for streaming video applications, where
mobile blocks can be sent together with the video frame data. The client
application is then also adapted by adding the necessary functionality---in the
client's source code---to handle the extra content of packets coming in from
the application server, and, if necessary, to respond by inserting responses in
outgoing packets.

\section{Mobile Data Blocks}
\label{sec:mobile_data}
When code fragments are made mobile, it suffices to replace all call
sites with stubs, and use a simple indirection step to either download
the code fragment, or to execute it immediately. In contrast, data blocks
can be accessed from any location in the program that can dereference a pointer
to the block. Due to the problem of
aliasing~\cite{alias}, precisely identifying all those locations for all
potentially useful mobile data blocks is impossible. Even if it would be
possible, adapting all code to ensure that a data block is downloaded before it
is accessed would introduce an unacceptably high overhead.

The solution is not to adapt the program locations where pointers are
potentially consumed, but instead to adapt the locations where pointers
to the data blocks are produced.

To produce an address of a statically allocated data section during the
execution of a program, three options exist. First, the address of some section
can be available in the statically allocated data of the program, i.e., in
another data section. Such cases are trivial to identify in object files,
as they are marked in the relocation information that linkers consume to relocate
such addresses.  The second option is that the
address of some data or data section is computed in a code fragment of the same
binary. Those cases can also be identified through the relocation information.
The third option is that the address of some section is produced or statically
stored in another binary (e.g., a library) that is loaded into the same
process. That case can only occur when at least one symbol in the section at
hand is exported from the binary containing the section. If no such symbol is
exported, it is impossible for the dynamic loader to let another binary relocate
a symbolic reference to the section.

In short, data sections linked into a binary become accessible if and only if
(i) a symbol residing in the section is exported from the linked binary,
or (ii) a relocatable address residing in the section is stored in another
section that is accessible, or (iii) a relocatable address residing in the
section is produced in code being executed.


These conditions for being accessible are already used by linkers. The GNU
linker option \texttt{--gc-sections}~\cite{binutils} lets it garbage collect all
inaccessible sections. Link-time program compaction techniques have pushed this
further by combining inaccessible section analysis with
whole-program unreachable code analysis~\cite{LCTES01}.
Our support for data mobility relies on the same principle: We limit mobility to
data blocks that (i) correspond to full data sections in the object files and
that (ii) become accessible only because their addresses are computed in code
that is marked to become mobile and possibly renewable. We exclude data sections
that become accessible because their addresses are stored in other data sections
or because they are exported.

The limitation to full data sections poses no problems for the forms of
renewability that will be discussed in Section~\ref{sec:applications}. Most
compilers offer a compilation flag \texttt{-fdata-sections} to store statically
allocated variables in a separate data section each. So the granularity for
making data mobile is that of individual global variables. This suits our
purpose.

The second limitation poses no problems for the forms of renewability we
currently support either, because we only make data mobile in connection with mobile
code.
When a source code fragment is annotated with code mobility pragmas,
and the option of data mobility is enabled in the pragma, the link-time rewriter
automatically identifies all data sections that become accessible only through
addresses produced in that code fragment. Those data sections are then
made mobile together with the code fragment. Our data mobility can
hence be seen as code mobility where statically allocated data ``owned'' by a mobile
code fragment becomes part of its mobile block. In Figure~\ref{fig:arch}, this
is visualized with arrows from mobile code blocks to mobile data
blocks in the heap memory region of the client-side application. Remember, those
arrows do not indicate that only the mobile code blocks can access the
data. They only indicate that the mobile code blocks contain the code fragments
that generate pointers to the mobile data blocks in the program state as the
mobile block is executed, thus making the mobile data blocks accessible to other
code fragments.

Because the Binder and the injected stubs ensure that each mobile code fragment
is downloaded before it is executed, and because they download the mobile data
together with the code that can produce the data's address, they also ensure
that mobile data is downloaded before any pointer to it is generated or used to
access the data.

\section{Tool Flow Support}
\label{sec:toolflow}

\begin{figure}
\centering
\includegraphics[width=0.70\columnwidth]{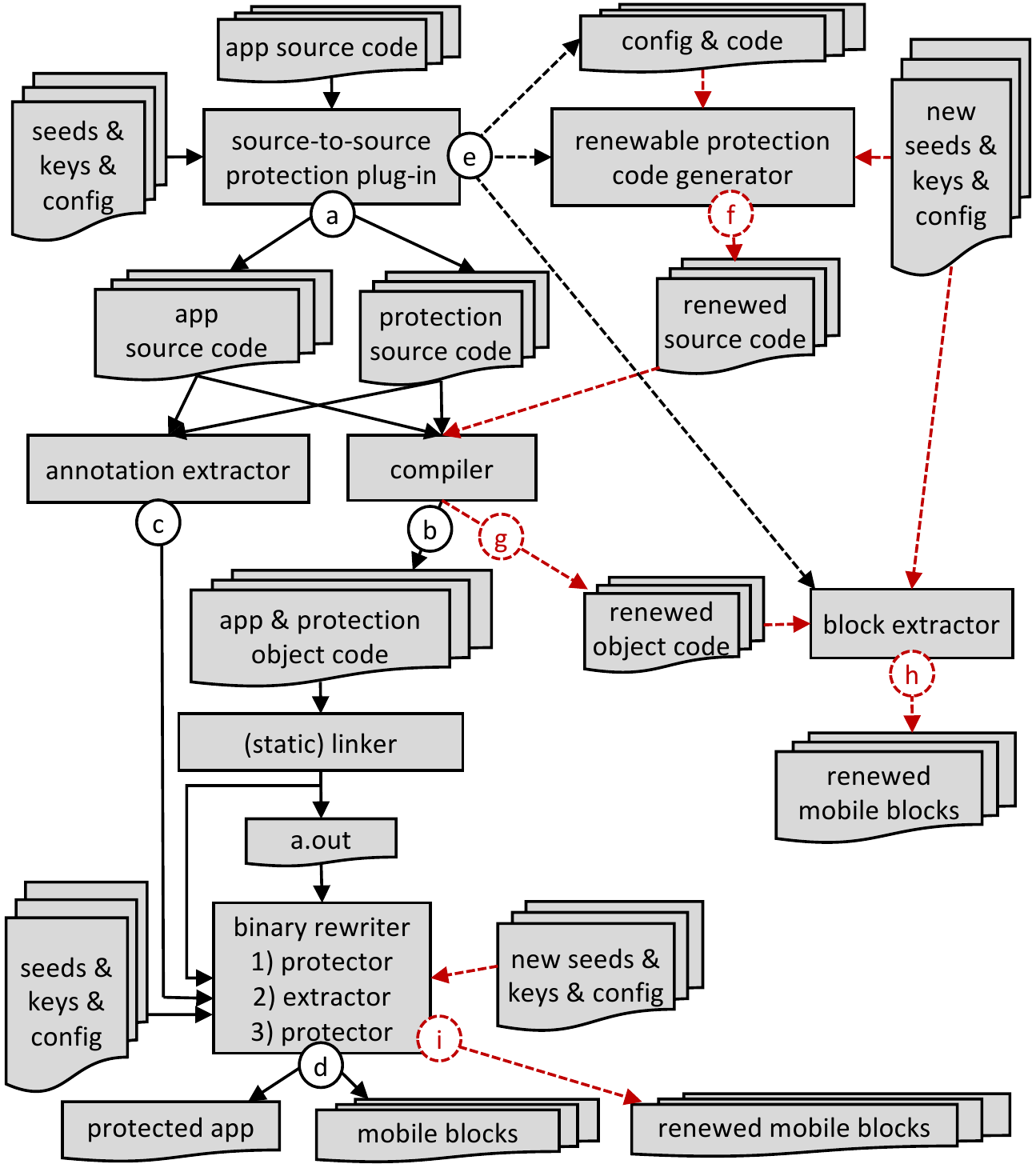}
\Description{The tool flow for generating mobile blocks. The normal tool flow is
  extended with generators (and related data) to allow for the regeneration of
  renewable blocks.}
\caption{\changed{Compiler tool flow for generation of protected application and renewed mobile blocks.}}
\label{fig:toolflow}
\end{figure}

Figure~\ref{fig:toolflow} depicts the tool flow that integrates
the renewability framework with protections. Full black arrows denote the
compiler and protection tool flow of code and data that includes basic code
mobility~\cite{cm} but
without renewability. Dashed black arrows denote the generation of renewable
code generators. This code generator generation process was added to the
existing tool chain for supporting renewability. Dashed red arrows visualize
the flow of code and data when renewed mobile blocks are generated, either a
priori or on demand.

\changed{Up-front, we want to clarify that the depicted tool flow extensions in support of different forms of renewability are not are not fundamentally new concepts. They are instantiations of known mechanisms to generate software diversity which can be done in many different phases of the software development life cycle as described in literature~\cite{Larsen14}. We simply developed instantiations that fit our overall tool flow design and the goal of ensuring composability of renewable protections with many other protections.}
 
\subsection{Existing Static Protections and Mobility Tool Flow}
The existing tool flow supports the insertion of software protections in three
phases.
First our tool chain contains a number of
source-to-source protection plug-ins. These take seeds, keys, and other
configuration parameters as input, together with the application source code to be
protected. In step \circled{a}, the plug-ins produce transformed, (partially)
protected application source code, as well as protection source code that
implements additional protection functionality to be injected into the
application. \changed{Examples of the latter functionality include functions that implement hashing for code guards and initialization routines for certain protections such as routines that set up the dynamic graphs used to which opaque predicate computations in the transformed application source code refer.}

The operation of the plug-ins is based on source code annotations
such as pragmas and attributes. These annotations allow one to mark the code
fragments that need to be protected and to specify the protections to be
deployed, their parameters and configuration. 
Figure~\ref{fig:toolflow} only depicts one source-to-source plug-in, but any
number of them can be chained in practice~\cite{aspire_toolflow}.

Both sets of source code produced by the source-level
plug-ins chain are then fed to a compiler to produce object files in step
\circled{b}. The compiler can optionally inject additional protections. In our
prototype, this is not the case, as we use a standard LLVM to compile Linux and
Android binaries. \changed{However, diversifying~\cite{multicompiler} and
obfuscating~\cite{junod2015obfuscator} forks of LLVM could be used just as well.}

From the source code files fed to the compiler, the remaining annotations and
their line numbers are extracted by an annotation extractor in step
\circled{c}. After the object files have been linked, both the object files,
the linked binary \texttt{a.out}, and the extracted annotations are fed to a
binary rewriter, together with additional seeds, keys, and configuration info.
The binary rewriter deploys binary-code-level protections, extracts blocks to
make them mobile, and applies further protections, both on the extracted code
and on the remaining, static code. In step \circled{d} the rewriter produces
the fully protected application as well as an initial set of mobile blocks as
specified by the code mobility annotations extracted from the source code. The
binary rewriter maps source code annotations onto binary code fragments  by
means of the extracted source line numbers and the line number information
present in the object files.

\subsection{Renewable Code Generator Generation}
The existing static tool flow is extended in several ways to enable renewability
for both protection code and original application code. First, the
spec of the source code annotations is extended to support
renewability. The tool chain documentation provides a full spec of the
annotations~\cite{aspire_toolflow}.

Secondly, source-to-source plug-ins are extended to produce not only the initial
code version, but also the necessary code and data for generating additional
code versions later on. See \circled{e} in Figure~\ref{fig:toolflow}. \changed{Per protection that can be made renewable, three
components are added}. First, a
\emph{renewable protection code generator} is needed, and its \emph{code and
  configuration} inputs need to be stored persistently. The generator is a
version of the plug-in that can be invoked separately, with new seeds, keys, and
other configuration parameters to generate different code versions. Its code and
configuration input contains a partial copy of the original source code and annotation input of
the plug-in.  This generator can be
application-specific, in which case it is produced or at least customized on the
fly by the plug-in during the source-to-source protection, but it can also be a
pre-installed tool.

To inject the renewed source code generated by the generator and compiled by the
existing compiler into actual mobile blocks, a \emph{block extractor} is
needed. This can also be application-specific or pre-installed. It knows, for
the form of protection supported by the plug-in, how to extract binary code
fragments and data sections from object files, which can trivially be done with
standard GNU binutils tools, and how to create new mobile block versions out of
them to be stored in the Diversified Block DB.

\subsection{Renewable Code Generation}
With the presented tool flow, renewed versions of code and data
blocks can be generated. For source-level protections, the generators are
invoked on their input codes and configurations, albeit of course with new,
different seeds, keys, and parameters. The result of this step \dcircled{f} is
renewed source code, either of the original application or of some
protection. This renewed source code is then compiled to produce renewed object
files in step \dcircled{g}, after which the block extractor extracts and
assembles renewed mobile blocks in step \dcircled{h}.

For binary-level protections, our prototype of the renewal process
re-runs the binary rewriter on its inputs with new seeds,
keys, and configuration parameters. The binary rewriter then produces renewed
mobile blocks in step \dcircled{i}. This is not very efficient,
as each invocation of the rewriter re-executes all the
binary-level processing passes, including passes on code fragments
that do not become mobile. With some engineering, this can
definitely be made more efficient.

\subsection{Discussion}
Neither the framework architecture nor the tool flow are limited to application
executables. As is, they can also be deployed to protect dynamically linked
libraries. In Figure~\ref{fig:toolflow}, both a.out and the protected app in
that flow can in fact be libraries such as libmylib.a.

In our proof-of-concept implementation resulting from the ASPIRE project, all
the necessary client-side components (Communication Logic, Downloader, Binder,
...) are linked statically into either an executable or into a dynamically
linked library, and non-exported symbols are stripped. This means that, e.g., the Binder cannot be identified by means
of symbol information, and also that its code is mingled and protected with the
original application or library code. This design choice does imply that when
multiple dynamically linked libraries protected with renewable protections are
loaded into the same application process, those components will be loaded and
executed multiple times, possibly even in parallel. To avoid this overhead, one
could opt to put the client-side components in a separate dynamically linked
library, of which only one copy then needs to be loaded into a process.

That would lower the level of protection, however, as all the
interfaces to those components are then exposed in the libraries' exported and
imported symbols. Furthermore, in that case a MATE attacker would only have to
attack one version of those components to defeat all
renewability in the process. When there is one copy per library, all of which
can be protected with different, independent anti-tampering and
anti-reverse-engineering protections, such as different forms of obfuscations,
remote attestation, and renewability, an attacker will have to invest much more
work.

Secondly, putting the components in an external library would imply that the
single version of each component that is then loaded into the running process
has to perform the renewability bookkeeping of multiple libraries that were
possibly compiled and protected completely independently from each other. This
would make those components much more complex, and it would significantly impact
important aspects of the software development life cycle. For example, it would
imply that only libraries protected with compatible forms of the renewability
support can be loaded together into a process. This would make it practically
infeasible to load protected libraries from independent vendors into the same
process, which would result in a DLL Hell as existed on Windows in the
past. In the current design, by contrast, every loaded library and the
renewability components in it are oblivious to the fact that other protected
libraries with renewability components are running in the same process. They can
even connect to different servers. This is obviously useful: It is
not unimaginable that vendors of different libraries (e.g., libraries that
implement vendor-specific DRM plug-ins for Android's media and DRM frameworks)
only trust their own servers.

Also on the server, each of the running libraries are treated in isolation. Even
if multiple libraries running in some client process connect to the same
server, that server does not know that its incoming requests are
originating from the same running process. This greatly eases the design and
development of the server functionality.

Of course, this design choice limits the flexibility of the server decision
processes. Currently, there is no global coordination between the renewability
services serving the multiple libraries that may be running in the same process,
and that hence may be undergoing the same attack. As future work, we plan to
investigate whether such coordination can be supported effectively and
efficiently.

\section{Mitigations Against Concrete Attacks}
\label{sec:applications}

The renewability framework supports a range of renewable
software protections that mitigate MATE attack steps.

\subsection{Syntactically Diversified Mobile Code}
\label{app:syntax}
Dynamic analysis is a common method for reverse engineering. 
It can be done manually, e.g., with a single-step debugger,
or it can be automated, e.g., by collecting trace information with an emulator. 
It can also be semi-automated, e.g., by writing small
debugger scripts that steer the program execution up to the specific point of
interest by means of breakpoints and watches, at which point manual single
stepping can start to collect additional information. Such scripts are often
developed iteratively: Each time more information is obtained, the scripts are
adapted to zoom in on the next piece of useful information on the attackers' path.
All of these approaches commonly involve multiple runs of the same program. This also
happens in, e.g., delta-debugging-like attacks, in which the difference in
program execution behavior on different inputs is
analyzed~\cite{artho2011iterative}, and it obviously also happens in fuzzing
attacks~\cite{sutton2007fuzzing}. Such attacks require repeatability, and become
harder if the code fragments that are revisited differ from one run to the
other.

This can be achieved by syntactically diversifying the code in renewed mobile
blocks. Rather than creating one version of a mobile block, multiple
semantically equivalent but syntactically different versions can be created and
delivered.

Our prototype tool flow creates versions by stochastically applying 
obfuscations (opaque predicates, branch functions, and control
flow flattening) and code layout randomization on the extracted code fragments. By initializing a pseudo-random
number generator with varying seeds, versions can be
generated that feature varying control
flow graphs and code layouts~\cite{IEEESP13}. This makes it
significantly harder, e.g., for attackers to automate the setting of breakpoints in
their scripts. It also makes it harder to compare multiple traces in collusion
attacks. 

The applied obfuscations have previously demonstrated their effectiveness in the
context of collusion attacks that rely on program
diffing~\cite{IEEESP13}, where they prevent diffing tools to automate the
identification of the corresponding code fragments in two syntactically
different versions of the same software. We therefore conjecture that it will be
non-trivial for an attacker to automatically overcome the protection provided by
syntactically diversified mobile code.


\subsection{Semantically Diversified Mobile Code}
\label{app:seman}
Syntactical diversification does not hamper all attack tools. For example,
pointer chaining tools (e.g., Cheat Engine - \url{https://www.cheatengine.org/})
can still find relevant data in randomized memory layouts during repeated
executions. In a first run of a program, the attacker then identifies the
relevant data in the process memory space manually. The tool then collects the
pointer chains to the identified data. These chains are lists of offsets. For
example, the transparency value of a wall in a shooter game might be located at
the end of the chain \texttt{*(*(*(frame\_pointer\_main+24)+4)+8)}, which does
not depend on the code syntax or layout, or on the data layout as affected by
address space layout randomization. Cheaters might want to make the walls
transparent to see their adversaries through them. For such chains to become
invalid for repeated executions, the order of fields in C/C++ structs and
classes needs to be diversified, the location where data is stored in stack
frames needs to be diversified, the order in which parameters are passed to
functions needs to be diversified, etc.

We hence need diversification that also changes the semantics of individual
fragments, i.e., the relation between the process state before and after
executing them. For example, when the fields in a C struct are reordered, code
fragments writing to the fields will write to different offsets in allocated
memory blocks, and hence implement different semantics.

Deploying such diversifications is more difficult, however, because they have a
more global impact on the generated code. If the order of fields in a struct is
altered, all code in the binary that accesses the struct will change as well, in
a consistent manner: the same change in offset will occur all over the program.
Likewise, if the signature of a procedure is altered, e.g., by reordering its
parameters, the procedure's code body will change, but so will the code of all
its callers. When aggressive compiler optimizations are used, those initial
changes can result in ripple effects throughout the binary code of all directly
or indirectly affected functions. In general, almost all data and data flow
obfuscations or diversification techniques~\cite{collberg2009book} have more
global effects on the generated code. To support such forms of diversification,
a client that is served multiple diversified code fragments by a renewability
server can only execute correctly if all of the fragments served during a single
run implement and assume consistent semantics.

Our approach supports this, because the server can partition the diversified
mobile blocks into consistent groups: The Renewability Manager can be informed
which versions of the mobile blocks in the database feature consistent
semantics, which do not, and which ones are independent.
Simple server-side bookkeeping can then ensure that whenever some
block is requested, it only delivers blocks that are independent of or
consistent with previously delivered blocks.

Of course, the use of this form of semantic fragment-level diversity restricts
the freedom of the renewability policies to replace fragments within a single
execution of a program: Once data values or the layout of data in the client
program's address space have been produced by a certain first code version, all
code executed later during that execution has to be consistent with that first
code version.

Still, the use of attack tools and reuse of attack scripts over multiple runs
of an application can be significantly hampered by this form of protection. In
particular, it will decrease the effectiveness of pointer chaining tools.

To support semantic renewability, we extended the basic tool flow of
Figure~\ref{fig:toolflow} somewhat. For a prototype implementation that changes
the layout and order of fields in structs and the parameter order of functions,
we rely on a source-to-source protection plug-in to generate the
diversified code. To enable the identification of all binary code fragments that
undergo relevant changes as a direct result of the source-level diversification
or as a result of ripple effects through compiler optimizations, multiple
approaches can be envisioned.

In our approach, we do not want to restrict or alter the used compiler. Instead,
in line with common industrial software development life cycle requirements, we
want to keep treating the used compiler as a black box. Then two options remain.
The first, more conservative option, is to track references to diversified
function signatures and structs in the rewritten source code, to mark any
function that directly or indirectly touches (directly or indirectly) upon such
references as dependent on the deployed diversification, and to enforce separate
compilation of each function in the compiler such that compiler optimization
ripple effects are bound to individual functions. This strategy will
conservatively over-approximate all functions or code regions that might be
affected by the deployed diversification, which allows us to make all of those
mobile and renewable. 

A more accurate identification of the altered binary code fragments, i.e., the
ones that need to be made mobile and renewable as a result of source-level
diversification and potential ripple effects, can be achieved through binary
diffing. To enable this diffing in our tool flow, we generate all diversified
versions up-front. We also compile all of them up-front. We then run a binary
differ that compares the compiled binaries, and identifies the precise
differences between the compiled versions. The binary code regions embodying
those differences are then marked to be made mobile and renewable in the binary
rewriter. Only after all versions are compared and all necessary regions are
identified do we run the binary rewriter to extract the necessary blocks from
all program versions. The remaining rewritten binary in which these blocks
are removed, is then identical across
all versions, as it consists of the binary code fragments that did not differ at
all across the different versions, whereas the mobile blocks contain all the
differing code.

This black-box approach offers the advantage of not needing any change to the
used third-party compiler and linker, or to the internal operation of the binary
rewriter. The developer of the source-to-source protection plug-in that
implements this semantic diversity hence does not need to invest any effort in
learning all the ripple effects that those three complex tools might induce as a
consequence of his source code transformations. The diffing tool automatically
exposes all code impacted by the ripple effects. We implemented our own
Clang-based source-to-source rewriter, but our approach easily allows for other
(already existing) source-to-source rewriters.

It is important to note that the semantic diversification does not need to be
limited to individual code fragments. If appropriate, it can easily be extended
to externally visible changes to the semantics of the whole program as well. For
example, in some cases it might be useful to renew the semantics of code
fragments that prepare a payload to be sent to the original application server
(see Figure~\ref{fig:arch}) or that consume a payload received from the
application server. Formally, this changes the semantics of the whole client
program, but if this is coordinated with the semantics implemented on the
application server, this can be perfectly fine, and happen transparently to the
end user of the software.

\subsection{Dynamic and Time-Limited White-Box Cryptography}
\label{app:wbc}
White-box cryptography (WBC) is a technique for protecting the confidentiality
of cryptographic keys in software~\cite{Chow02wbAES,WBC}. The literature mostly
focuses on fixed-key implementations, where the key is hard-coded in the
software. Rather than including a key as a constant input to a standard
implementation of a cryptographic primitive, which is trivial to attack in a
MATE scenario, a custom version of the primitive is included in the software,
which hard-codes the key in a way that it cannot be extracted (easily), e.g., by
encoding it in large randomized tables or code structures.

Fixed-key implementations are acceptable for some use cases such as hard-coding
global bootstrap keys. However, for many industrial use cases keys need to be
updatable. For example, for personalizing software with application-unique keys
or for installing service-dependent keys, cryptographic implementations can
ideally be instantiated with keys at run time~\cite{WhiboxIndustrial}. While
there is almost no literature on this, several companies are selling such
\emph{dynamic-key white-box} implementations; there is no publicly available
information on they are built. One possible approach would be to build
special-purpose white-box implementations which receive a protected version of
the key as input. 
would trivially expose the key.  The protection of the keys then needs to be
integrated in the application design, and additional routines such as
preprocessing the protected key and key schedule algorithms need to be
integrated: This introduces a lot of additional complexity and can have a
considerable impact on performance and code size. Another approach is to update
existing fixed-key implementations at run time. In the most common white-box
implementations such as that of Chow et al.~\cite{Chow02wbAES}, key material is
embedded in look-up tables. It suffices to update these tables in order to
change the key, so the technique of mobile data blocks can be applied to achieve
dynamic white-box implementations. Other white-box techniques do not solely
depend on look-up tables~\cite{Billet2003,Bringer2006}, but also encode the key
in complex code structures. Updating the key then implies updating the
code. This is also supported out-of-the-box with our renewability framework,
\changed{as the code to be updated can trivially be annotated to be made
  mobile}. In summary, our framework offers all that is required to evolve from
static key WBC to dynamic key WBC.


Still, designing secure WBC implementations, whether static or dynamic, remains
a challenge. All currently proposed designs have been broken, and recent
proposals that are submitted to the ECRYPT White-Box Cryptography Competition
(the WhibOx Contest)~\cite{WBChallenge} are challenged in a matter of hours or
days. Therefore, rather than focusing on designing implementations that give
long-term security guarantees (and will probably be very slow and large) an
alternative approach is to focus on more efficient but less secure
implementations that are renewed at high frequencies. We denote these as
\emph{time-limited white-box} implementations. Such WBC implementations can
protect short-lived session keys or temporary access tokens with acceptable
performance. With those implementations, it are then not the keys that need to
be rotated frequently, but the WBC implementations that embed the keys. This
rotation is readily supported by our renewability framework.

Combining this form of renewability with the already discussed forms of
diversification can then help to achieve longer-term protection as well, namely
by ensuring that the rotated implementations differ in more respects than simply
embedding different keys, thus hampering attackers in reusing simple attack
scripts. 


\subsection{Diversified Static-To-Procedural Conversion}
Static data such as strings can serve as hooks in many attacks. To protect
these against static inspection, static-to-procedural
conversion~\cite{collberg2009book} replaces the static data by invocations to
injected procedures that compute the data on the fly. If dynamic attacks are
then also made harder, e.g., by combining this protection with
anti-debugging~\cite{SSPREW16}, strong protection can be obtained. With
renewability, the level of protection can be increased even further: If the
code that computes the data changes between every run of a program, the
attacker will have to adapt and re-execute his attack script to extract the
data he is after whenever a new version is downloaded.

This form of renewability is readily supported: It suffices to let the
source-to-source protector generate randomized procedures to replace static
data, and to annotate these to have them extracted by the binary rewriter.

\changed{In a sense, this form of renewability is situated somewhere in between
  syntactic and semantic diversification: the overall semantics of the renewed
  code blocks stays the same, as they produce exactly the same constant data,
  but they may do so using widely varying algorithms. Implementation-wise, it is
  similar to WBC in the sense that the procedures generated at
  build-and-protection-time simply need to be annotated to be made mobile, and
  the generator needs to be invokable at deployment time to generate different
  versions.}

\subsection{Diversified Instruction Set Randomization}
\label{app:ISR}
A popular form of obfuscation is to translate an application or part thereof is
to some virtual, randomized bytecode instruction set architecture (ISA). At run
time, the bytecode is emulated. Popular tools that implement this form of
emulation-based obfuscation are Code Virtualizer~\cite{CV},
EXECryptor~\cite{exec}, Themida~\cite{themida}, and VMProtect~\cite{vmprotect}.
Unlike native code formats---which are well-documented by processor
manufacturers---the randomized ISA is not documented. It is also diversified
for each protected program to reduce the learnability for attackers.

Custom bytecode can be made mobile when it is read-only data where the set of code locations
that refer to it is clear and limited. In the ASPIRE tool chain, security-sensitive chunks of native code are translated into bytecode
chunks~\cite{wicsa2016reference,D1.04}.
The original chunks are replaced by stubs that invoke the emulator, passing it
a pointer to the data representing the bytecode. This scheme fits our mobile
data block support perfectly: The stub becomes mobile code, and the
bytecode---to which only the stub produces a pointer---becomes a mobile data
block attached to that mobile code.

Bytecode renewability can then be achieved by combining diversified mobile
bytecode with semantic diversification of the emulator. Both
the semantics and the syntax of the bytecode can then vary over time; for each
execution a corresponding interpreter and bytecodes are delivered. 

\subsection{Evolving Protections}
\label{app:evolving}

The proposed tool flow and architecture pose no limits on the sizes of the
mobile blocks. In particular, renewed blocks don't need to have the same sizes.
This helps in supporting gradually evolving
renewability, e.g., where over time more complex forms of protections are
delivered to client applications as those protections become available in
response to detected attacks.

For example, when more advanced attacks on WBC crypto schemes become
available over time that reduce the search space for brute-force attacks, or
when faster brute-force methods become available, more complex versions of the
white-box algorithms can be delivered to the client applications to catch up
with the attacker's capabilities. In many WBC schemes, this can be achieved with
bigger tables that embed the secret keys.

\changed{Another example is that of an evolving license check. Many applications
  are distributed free of charge, but a license key needs to be acquired through
  an online transaction in order to run the application or to make the full
  functionality available. Often, a legitimately acquired license key will work
  on multiple versions, including software updates released after the original
  transaction. Typically, individual buyers get individual, fingerprinted
  license keys to enable the vendor to trace illegitimate redistribution of
  keys. By contrast, the code that checks their validity is the same for all
  buyers of a specific version of the software. Once that code is
  reverse-engineered, it becomes straightforward for crackers to generate license
  keys to sell on the black market. However, the business model of those
  crackers can be undermined, and the trust placed in them by their customers
  can be broken, by renewing and extending the validity checks in each released
  version. Whereas the vendor can know all validity checks up-front, even those
  that will only be included in future releases, the cracker only knows the
  checks in the reverse-engineered version. So whereas the vendor can easily
  generate forward-compatible keys, the cracker cannot. This obviously has a big
  negative impact on the value of the keys they generate. Traditional
  implementations of this protection scheme, in which the checks are embedded
  statically in released versions are obviously limited in terms of the
  frequency with which the checks can be updated. With renewability as delivered by
  our framework, it is easy to overcome this limitation.}

To support such evolving protections with our framework, the only aspect that
needs to remain constant from one mobile block to another is the
binary-level interface of each renewed code block, i.e., the way data is passed
to and from the mobile blocks to static code, and in between mobile blocks: the
registers used, the stack frame layout, ...


For source-level forms of diversification and renewability, this requirement of
constant binary-level interfaces can in practice only be achieved when the
mobile code blocks correspond to units of which the compiler cannot alter the
binary-level interface at will.
This is the case for whole functions or methods, because compilers are bound to
calling conventions.
Functions and methods are often also the ``units'' in which developers implement
functionality, be it protection, library, or application functionality. So in
practice, the limitation to renew only whole functions does not impose overly
strict restrictions on the ability to let the deployed protection components
vary over time. \changed{Importantly, renewing whole functions is already
  supported out-of-the-box by our proposed (and prototyped) tool flow: It uses
  the exact same infrastructure that is used to update whole WBC functions.}

Besides the potential to respond to advances in the attacker's toolbox, this
ability to vary the deployed protection offers two major advantages. First, it
can help in reducing the time to market. Selecting the optimal combination of
software protections is a cumbersome, difficult, time-consuming task.
The framework's capability to vary protections
over time allows vendors to release weaker protected versions early, and to
upgrade the protection seamlessly (without the user being disturbed) after the
initial release. Second, the ability to vary the deployed protections over time
can be used to find a better balance between their strength and overhead.  A
good example is code integrity verification by means of Remote Attestation (RA)
based on code guards. Code guards are basically hashing functions that compute
hashes over the code being executed. With RA, a server requests such a code
guard to be executed on some code region, and checks whether the received hash
value is the expected one. If not, this is a signal that the code has been
tampered with.
Different RA and code guard designs come with different degrees of overhead. To
keep the overhead acceptable, all schemes leave some freedom to the attacker to
tamper and remain undetected.
When the deployed scheme varies over time, however, as supported by our
approach, the attacker has to take into account all possible schemes to remain
undetected for a longer period of time. As already discussed before, attacks
often involve multiple executions of a program, so this period typically spans
multiple executions. During any (tampered) run then, the attacker has to be
cautious and assume no freedom, as if all anti-tampering schemes were being
deployed together. At any point in time, however, only one or a couple of
schemes are actually deployed. A regular user thus only experiences the
overhead of a limited number of them. In the ASPIRE project, we experimented
with renewing code guard implementations that, e.g., vary the pseudo-random
walk over the code fragments they hash. By making it unpredictable for an
attacker which instructions will be visited, it suffices to hash only a limited
number of instructions during any invocation of a guard.




\section{Experimental Evaluation}
\label{sec:evaluation}

\subsection{Target Platform of Prototype Implementation}
Our prototype targets ARMv7 client platforms. Our client hardware consists of several
developer boards, on which we ran Linux
3.15 and Android 4.3+4.4. For Linux, we used a Panda Board featuring a single-core Texas
Instruments OMAP4 processor, an Arndale Board featuring a double-core Samsung
Exynos processor, and a Boundary Devices Nitrogen6X/SABRE Lite Board featuring a 1GHz
quad-core ARM Cortex A9 with 1 GByte of DRAM. The latter was also used for running all Android
benchmark versions, and for running the measurement experiments reported
below. On the server side we set up a VirtualBox VM running a 64-bit Debian
Linux, 2 GBytes of RAM and a Gbit NIC adapter. This VM ran on an Intel Xeon
E3-1270 CPU 3.50GHz with 16 GBytes of RAM. We used GCC 4.8.1, LLVM 3.4, and GNU
binutils 2.23 for the client, for which we compiled code with {\small\texttt{-Os
    -march=armv7-a -marm -mfloat-abi=softfp -mfpu=neon -msoft-float}}. On the
server we used GCC 4.8.1 and binutils 2.23 to build our components. The
Mobility Server and Renewability Manager were compiled with
{\small\texttt{-O3}} and {\small\texttt{-Os -fpic}} respectively. \changed{Our techniques to protect native code do not depend on features of the mentioned, relatively older versions of system software. Porting our prototype implementation forward to newer versions requires engineering work, however, because small patches are needed to the used compilers and assemblers to have them generate enough symbol information and non-relaxed relocation information for the link-time rewriter. While that work is certainly doable, porting the whole Android use cases that we introduce below, including all their Java code that is not a target of our techniques, requires a major effort that we cannot afford, and that does not have any impact on the presented work or results.}

\changed{To the best of our knowledge, the presented renewability techniques and all protections supported by our prototype tools can also be implemented for other platforms such as Apple's MacOS or Microsoft Windows, and for other architectures such as 32-bit and 64-bit Intel architectures. We have experience writing system software for those platforms and architectures, and we are confident that there are no fundamental obstacles. However, porting the implementation, in particular the binary-rewriting part and the dependence of certain protections on platform APIs, would require a huge engineering effort.}

\changed{Technically, Apple's iOS appears compatible with the proposed techniques, in the sense that the OS and related system software offer the necessary functionality. However, policies such as requiring that all executed code is part of the original binaries distributed via app stores, can obviously form a hurdle to start deploying the protections. Our research focuses on technical aspects, policy issues are out of the scope of this paper.}

\subsection{Correctness and Applicability Validation}

\begin{table*}[t]
\caption{\changed{Features of four evaluation use cases.}}
\centering
\resizebox{\textwidth}{!}{
\begin{tabular}{|l|c|c|c|c|c|}
\hline
use case & developer    & SLoC  & 3rd-party & assets & deployed forms of renewability\\
         &              &       & libraries            &   &  \\
\hline
\hline
DRM library & Nagravision  & 306.2k & OpenSSL & crypto & mobile code,\\
    &                    &         &         & keys            &  diversified \& dynamic (time-limited) WBC (Section~\ref{app:wbc})\\
\hline
software license & SafeNet & 55.4k & tomcrypt, & keys,    & syntactically diversified mobile code (Section~\ref{app:syntax}), \\
manager library         & Germany             &        & tommath   & code IP & diversified instruction set randomization (Section~\ref{app:ISR}) \\
\hline
bzip2 app        & Julian Seward    & 5.8k &   &   & syntactically diversified mobile code (Section~\ref{app:syntax})\\
                 & (open source)    &      & - & - & semantically diversified mobile code (Section~\ref{app:seman}) \\
                 &                  &      &   &   & \changed{evolving code guards (Section~\ref{app:evolving}})\\
\hline
WBC crypto    & Du\v{s}an Klinec & 6.3k & - & crypto & diversified WBC (Section~\ref{app:wbc})       \\
app              & (open source)    &       &   & keys   & \\
\hline
\end{tabular}
}
\label{table:usecases}
\end{table*}

\subsubsection{Industrial Android Use Cases}
First, the correctness and applicability of our framework were validated and
evaluated by deploying various forms of renewability on two industrial use cases
that were developed independently by two market leader companies using different
development approaches, software architectures, and build systems. Each use case
consists of a shared library of sufficient complexity to represent real software
products. Each of them embeds security-sensitive assets representative of the
assets in the companies' real products. We chose the code and data fragments to
make mobile and renewable together with the application architects, the
application developers, and security architects from the companies.

The first use case consists of two plug-ins, written in C and C++ at Nagravision
S.A., for the Android media framework and the Android DRM framework. These
plug-ins, in the form of dynamically linked libraries, are necessary to access
encrypted movies. A video app programmed in Java is used as a GUI to watch the
videos. This app communicates with the mediaserver and DRM server processes
(i.e., daemons) of Android, informing the daemons which vendor's plug-ins they
require. On demand, the daemons then load the library plug-ins. %
Concretely, these servers are the \texttt{mediaserver} and
In our research, we observed several features that make this use case a perfect
stress test. The multi-threaded mediaserver launches and kills threads all the
time. The plug-in libraries are loaded and unloaded frequently, sometimes the
unloading being initiated even before the initialization of the library is
finished. As soon as the process crashes, a new instance is launched. Sometimes
this allows the Java video player to continue functioning undisrupted, sometimes
it does not. These forms of behavior stress all client and server components.

The second use case is a software license manager that stores credentials, and
controls access to licensed content and functionality, e.g., through
time-limited and key-enabled licenses. This manager is programmed in C at
SafeNet Germany GmbH. It is a dynamically linked library that includes the JNI
interface, and is embedded in an Android app. This native library thus functions
as a license manager for a Java application. In this case, the Java application
is relatively simple: It is a riddle game of which the solutions are protected
by the license manager. To test our renewability support, this use case is also
interesting. In particular, the library is loaded into Android's Dalvik
execution environment, which features multiple threads (such as for the JIT
compiler, garbage collector, ...), and over which we have absolutely no
control~\cite{bornstein2008dalvik}. A command-line version of the riddle game,
programmed in C, is also available. It uses the same library (except the JNI
wrapper). On top of providing an easier target to debug on our Android developer
boards, this command-line version can also be compiled for Linux. This way, we
could also test our implementation on Linux.

Table~\ref{table:usecases} lists a number of features of the two use cases as
an indication of their representativeness of real-world software. The number of
source code lines includes all the mentioned third-party libraries that are
compiled and statically linked into the shared libraries to be protected. 
Whereas those linked-in libraries do not contain any assets, they operate on
assets such as keys, and their control flow hence needs to be protected against
reverse engineering as well.

Even though no additional protections are listed in Table~\ref{table:usecases},
we did actually combine many additional non-renewed protections with the listed
forms of renewability on the industrial use cases. This includes
anti-debugging~\cite{SSPREW16}, remote attestation~\cite{Viticchie16}, and code
and data obfuscation techniques~\cite{collberg2009book}. \changed{This way, the
  composability with other protections and correct functioning of the basic
  renewability components and of their deployments for specific forms of
  renewability as listed in Table~\ref{table:usecases}, were stress-tested
  extensively.}

\changed{Our testing effort included the following activities:
\begin{itemize}
\item checking logs produced by the tool flow to check that protections were
  deployed as foreseen,
\item checking the produced code for the presence of the protections and the
  artifacts resulting from their deployment,
\item running the protected use cases on the developer boards,
\item having professional red teams in (or associated with) the aforementioned
  companies perform penetration tests for several months on the protected use
  cases to validate the effectiveness of the protections against many different
  attack activities.
\end{itemize}
The first three activities were performed for all protections supported by our
prototype implementation. By contrast, the professional pen testing was only
performed on the non-renewed versions of the protections. The reason was the
ASPIRE project plan, in which the pen testing part of the validation work
package was executed in parallel with the final research development, which
included renewability. The practical results from those pen tests and the
broader validation effort in the project has been published in a public
deliverable~\cite{D106}. The knowledge acquired during those pen
tests regarding attacker activities and processes on programs protected with the
non-renewable versions has been systematized and published as
well~\cite{emse2019}. The follow-up and analysis of the pen tests allowed us to
pinpoint specific attacker activities that exploit weaknesses of the
non-renewable protections. A concrete example is the iterative refinement of an
attacker's tracing scripts to iteratively locate the most relevant code in an
execution trace. Internally, and after the ASPIRE project and the development of
the renewability had already finished, we then checked whether the renewable
forms of the protections effectively mitigated those pinpointed attack
steps. While we cannot give more concrete details because of confidentiality
agreements with respect to the professional pen tests, we can confirm that the
deployed protections indeed delivered the foreseen mitigations as discussed
throughout Section~\ref{sec:applications}.}

\subsubsection{Smaller Linux Use Cases}
The third row of Table~\ref{table:usecases} lists bzip2, the popular compression
tool. While this open source program does not contain any security-sensitive
assets, we used it \changed{to evaluate the correctness of our tool support for
  two additional applications of renewability, being syntactically and
  semantically diversified mobile code blocks on the one hand, and evolving
  protections on the other hand.}

For the former, we evaluated two semantic source-to-source diversifications:
struct field reordering and function parameter reordering. The correctness of
the semantic code diversification transformations was evaluated by compiling and
testing the diversified code as it was diversified with the source-to-source
plug-in. The correctness of the whole semantic code diversification setup was
evaluated by deploying the full extended tool flow, including the binary diffing
and mobile block extraction during binary rewriting, on multiple diversified
versions. For all of them, the exact same static binary with mobile blocks
extracted was obtained, and that binary was tested to execute correctly with any
compatible version of renewable blocks delivered to it.

Next, we investigated how the degree of semantic diversification (of
Section~\ref{app:seman}) influences the generated binaries and mobile blocks. In
our flow, a set of blocks that is mutually compatible originates from the same
diversified instance of the program. Any function that is diversified in any
specific instance needs to be made mobile in all instances in order for them to
be compatible with the same binary. Thus, increasing the number of diversified
program instances---from which the renewable sets of blocks originate---will
have an effect on the number of blocks that need to be made mobile, and on the
size of the remaining static binary. To gain some insights into these effects,
we experimented with bzip2 and function parameter reordering. Our tool flow has
a configuration parameter to specify the number of different versions that need
to be generated by diversifying the code of selected software components. We
varied this parameter value from 2 to 100. For each evaluated value, we did not
select any specific subset of functions for diversification, but instead allowed
the tools to randomly select any subset of functions in the whole program to
reorder their parameters. For each of the selected parameter values, we ran a
total of 20 differently random-seeded runs of our framework, and averaged the
measurements. We also measured the size of the .text section of the
undiversified bzip2 binary both with and without the extra support code that is
linked into the binary to support the code mobility functionality. This `base'
.text section consists of 94.9KB without support code, and 1116.6KB with; it is
thus clear that for this specific use case the mobility support code exceeds the
original application code by an order of magnitude.

Table~\ref{table:semantic-renewability} shows some results. For every number of
compatible programs, we measured the averages of: the number of functions found
to differ and thus made mobile, the total size of mobile blocks proportionate
to the base .text section, and the percentage of base .text still present in the
binary. It can be seen that the portion of the binary being made mobile
increases with the requested number of diversified versions, but that there is a
limit to this increase. There might be code that will never be impacted by the
specific diversifications used, and thus need never be made mobile (as a simple
example, leaf functions without parameters can never need to be made mobile with
this specific diversification transformation). Next to that, the transformations
used for code mobility both increase the size of the mobile code, and the size
of any code still left in the binary invoking the mobile code. Note that the two
fractions add up to more than 100\% because making fragments mobile involves the
injection of stubs and other small code snippets in the static binary, and
because the code in mobile blocks is enlarged as it is transformed to make it
offset-independent as discussed in Section~\ref{sec:architecture}.

\begin{table}[t]
\caption{Effects of increasing the number of diversified versions for function parameter reordering}
\small
\centering
\begin{tabularx}{\textwidth}{c|YYY}
\changed{ \textbf{\# versions}} & \changed{\textbf{\# mobile functions}} & \changed{\textbf{total mobile block size relative to .text size}} & \changed{\textbf{remaining fraction .text}}\\ \hline
2        & 3    & 8.0\%   &	93.6\%                                 \\
5        & 8    & 24.4\%	&	79.8\%                                 \\
10       & 15   & 37.5\%	&	68.8\%                                 \\
20       & 22   & 53.9\%	&	54.7\%                                 \\
50       & 31   & 77.7\%	&	33.1\%                                 \\
100      & 32   & 79.5\%	&	31.4\%
\end{tabularx}
\label{table:semantic-renewability}
\end{table}

\changed{As an instance of an evolving protections, we adapted the existing,
  initially non-renewable, offline code guard support in our tool flow. The
  protection of offline code guards protects the integrity of the code in two
  steps. First, invocations of attestation functions are injected into the
  program at selected program points. Those functions compute checksums in the
  form of hashes over parts of the application code in memory as specified by
  the developer by means of source code annotations. The regions are encoded in
  a data blob that the binary rewriter tool flow component injects into the
  protected binary. Secondly, invocations to verifier functions are injected at
  selected program points. These verifier functions check whether the computed
  checksums equal the expected values. If not, this implies that the code has
  been tampered with. In that case, the verifiers trigger an appropriate
  reaction such as aborting the program or corrupting the program state. The
  user of the protection tool flow has to select and provide the reaction code.


  We adapted the existing non-renewable prototype to make it renewable. Both
  attestators and verifiers can be renewed, allowing us to let the used
  attestation code evolve over time, e.g., to alter the order in which
  instructions in a code region are hashed or the used hashing function, and to
  alter the way the correctness of the result is checked. By making the
  attestators' and the verifiers' code renewable, their associated data blob
  becomes renewable as well. This allows us to let the parts of the program that
  are guarded evolve over time. To test the correctness of our implementation,
  including the ability to vary the attestators and the verifiers within a
  single run of the program, we configured the renewability policy to force a
  renewal in between different rounds of compression. Before different rounds
  are executed, different parts of the compression routines are attested and
  verified, as chosen by the renewability server. By means of the necessary
  logging functionality in the injected functions, we have been able to validate
  that the functionality works correctly and as intended.}

Finally, we tested diversified WBC on a small stand-alone WBC crypto app, of which some details are listed on the bottom row of Table~\ref{table:usecases}. While
we did so mainly to perform overhead measurements on which we report
later, they also contribute to the validation of the prototype implementation
and hence the practicality of the proposed approach.

\subsubsection{Conclusion}

In summary, the forms of renewability listed in the rightmost column of Table~\ref{table:usecases} have
been validated extensively on four use cases. Combined, our evaluation covers five applications from Section~\ref{sec:applications}. Most importantly, it covers both mobile and renewed code, and mobile and renewed data ---thus
covering all client functionality--- as well as most (and definitely all core)
server functionality. Finally, the evaluation successfully covers Android and
Linux platforms, and stand-alone application executables as well as dynamically linked
libraries.

\subsection{Performance Overhead}

In our previous work, we already analyzed the overhead of basic code mobility
when it is deployed over various wired and wireless networks with different
throughputs and latencies~\cite{cm}. The difference between basic code
mobility and renewability is the flushing and re-downloading of code after the
initial download. The impact thereof on performance obviously depends on the
frequency with which code needs to be flushed, as well as on the frequency with
which it needs to be downloaded. The flushing frequency is determined by the
enforced renewability policy. This hence varies from one usage scenario to
another, and even from one asset to another. The re-download frequency depends
on the flushing frequency, but also on the frequency with which the mobile code
and data is executed and accessed. As an extreme example, a code fragment that
is only executed when a new movie is launched in a media player, will need to
be downloaded at most once per movie, however fast it is flushed after that
execution. By contrast, a code fragment that is executed once or more per frame
in the movie will need to be reloaded at essentially the flushing frequency. %

The performance overhead of the proposed renewability protection will hence vary
wildly from one scenario to another. We therefore aim for providing the reader a
feeling for the range of overhead to expect, rather than for trying to argue
that the overhead is low enough. What is acceptable and what is not, depends on the
usage scenario at hand.

We did not measure the timing of the interactive industrial use cases. We can
confirm, however, that the overhead of the renewability did not
significantly impact the overall user experience of those apps. In the case of
the DRM library, downloading mobile code produces a slight additional delay when
a movie is started, but this delay is negligible compared to the delay caused by
having to download enough frames to fill the video buffer. The video playback
frame rate was not impacted by the renewable protections. The renewable
functionality of the license manager is downloaded when the software is launched,
and whenever functionality with custom licenses is accessed for the first
time. On those occasions, the downloading of code introduces a (barely
noticeable) delay that is deemed acceptable.

Our first quantitative performance analysis was carried out on the CPU-intensive
bzip2 program (\url{www.bzip2.org}). The experiment consisted of measuring
different properties of multiple runs of bzip2 over the controlled, standard
input consisting of the SPEC2006 training data (\url{www.spec.org}).
Experiments were carried out on three program versions, in which different sets
of functions were made renewable. For the first two versions, we collected
profile information with the GNU gprof tool~\cite{gprof}, and selected hot
functions of which the total execution time approximated respectively 20\% and
50\% of the total execution time of the program. The second set is not a
superset of the first one, but there is some partial overlap. In the third
version, all functions in the bzip2 program are made renewable. This corresponds
to 100\% of the total program execution time. It is hence clear that this
experiment is not meant to measure realistic overheads. Instead, the experiment
serves the purpose of a sensitivity analysis, demonstrating that the performance
overhead can be impacted by tuning the protection deployment, and that there is
a need to do so, because not doing so will often result in unacceptable amounts
of overhead.

With each version, we first set up a baseline
by collecting the execution time of a non-protected, vanilla application.  For each of the three renewability percentages, we then ran the
program for different renewability flushing time-outs of 1000, 2000, 3000, and
5000ms. For each mobile block, 600 different versions were generated a priori,
using syntactic code diversification techniques~\cite{IEEESP13}. On each
download request, the Renewability Server picks one of them randomly.

For each run we sampled the wall-clock execution time, the number of
transferred blocks, their total size in bytes, and the
CPU time consumed by the Renewability Manager on the server side.
Each experiment was repeated 20 times to collect data, in the remainder of this
section, we discuss and present averages over those 20 runs.

Table~\ref{table:execution-time} reports the average wall-clock times and the
overhead in that regard, as well as the network overhead in terms of numbers of downloaded blocks and the network
throughput. Table~\ref{table:server-side-overhead} reports the CPU
time consumption on the server. For reference and comparison,
Table~\ref{table:code-mobility-overhead} presents the overhead when the
different amounts of code in bzip2 are made mobile, but never flushed and
renewed, i.e., when they are downloaded only once. The tables confirm that the
overhead is directly related to both the renewal refresh rate and the hotness of
the code fragments being renewed.

\begin{table}[!t]
\caption{Client wall-clock execution times and network throughput of renewability on bzip2}
\small
\centering
\begin{tabularx}{\textwidth}{c|c|YYY|YYY}
\textbf{mobility}                   & \textbf{refresh time (s)} & \multicolumn{3}{c|}{\textbf{execution time (s)}} & \multicolumn{3}{c}{\textbf{transferred blocks}} \\ \hline
                                &                  & \multicolumn{1}{l}{Mean} & StDev & overhead  & \multicolumn{1}{l}{Mean}& \multicolumn{1}{c}{per sec.} & kb/s\\ \hline 
\textbf{0\%}                    & -                & 279                     & 0.3 \   & - \ \   & -\ \  & - \ \ \  & - \ \ \\ \hline
\multirow{4}{*}{\textbf{20\%}}  & 1                & 324                     & 1.6 \   & 16\% & 753                                   & 2.32 \   & 18.38\\
                                & 2                & 321                     & 1.9 \   & 15\% & 401                                   & 1.25 \   & 9.94 \\
                                & 3                & 319                     & 1.0 \   & 14\% & 276                                   & 0.86 \   & 6.93  \\
                                & 5                & 317                     & 1.0 \   & 14\% & 171                                   & 0.54 \   & 4.34  \\ \hline

\multirow{4}{*}{\textbf{50\%}}  & 1                & 487                     & 1.9 \   & 74\% & 3,885                                 & 7.97 \   & 12.77\\
                                & 2                & 475                     & 3.7 \   & 70\% & 1,953                                 & 4.11 \   & 6.83 \\
                                & 3                & 459                     & 1.1 \   & 64\% & 1,267                                 & 2.76 \   & 4.63  \\
                                & 5                & 456                     & 2.9 \   & 63\% & 793                                   & 1.74 \   & 2.90 \\ \hline

\multirow{4}{*}{\textbf{100\%}} & 1                & 647                     & 4.9 \   & 132\% & 9,818                                & 15.17 \  & 30.47 \\
                                & 2                & 591                     & 10.4 \  & 112\% & 5,236                                & 8.86 \   & 18.99 \\
                                & 3                & 565                     & 3.3 \   & 102\% & 3,498                                & 6.19 \   & 13.29 \\
                                & 5                & 552                     & 4.1 \   &  98\% & 2,127                                & 3.85 \   & 8.23\\ \hline

\end{tabularx}
\label{table:execution-time}
\end{table}

\begin{table}[!t]
\caption{Server CPU consumption for bzip2}
  \centering
  \small
\begin{tabularx}{0.6\textwidth}{c| *{4}{Y}}
\multirow{2}{*}{\textbf{mobility}} & \multicolumn{4}{c}{\textbf{renewability refresh time (s)}}	\\
\cmidrule(lr){2-5}
& \textbf{1} & \textbf{2} & \textbf{3} & \textbf{5}		\\\hline
20\%         				& 405 & 363 & 334 & 300 \\
50\%		 				& 923 & 621 & 544 & 439 \\
100\%         			& 1,006 \ \  & 860 & 669 & 565 \\
\end{tabularx}
\label{table:server-side-overhead}
\end{table}

\begin{table}[!t]
\caption{Baseline overhead of code mobility on bzip2}
  \centering
  \small
  \begin{tabularx}{0.7\textwidth}{lc *{3}{Y}}
& & 							\multicolumn{3}{c}{\textbf{mobility}} \\
\cmidrule(lr){3-5}
& &								\textbf{20\%} & \textbf{50\%} & \textbf{100\%} 	\\\hline
\multirow{3}{*}{\textbf{Client exec time (s)}}			& \textbf{Mean} 				& 282   & 299   & 313       		\\
& \textbf{StDev} 				& 211   	& 135		& 136					\\
& \textbf{overhead} 			& 1.1\%   & 7.1\%   & 12.0\%  	     	\\
\textbf{transferred blocks}								& 				 				& 4         & 22        & 55       		\\
\textbf{blocks/s}              							& 								& 0.01      & 0.07      & 0.18         \\
\textbf{network throughput (kb/s)}     					& 								& 0.08      & 0.06      & 0.18
\end{tabularx}
\label{table:code-mobility-overhead}
\end{table}

Comparing the server overhead to the client execution times, we
observe that for this program and hardware, the server CPU load varies
between 0.1\% and 0.2\% of the client load. Scalability on the server
is hence another factor to be considered when deciding on the
use of renewability, on the fragments to be made renewable, and on the renewal
policy enforced by the server. The same obviously holds for scalability of the
network capacity.



A similar experiment with a C++ WBC crypto
application was based on Du\v{s}an Klinec's implementation of the Chow
WBC scheme without external encodings~\cite{Chow02wbAES}, available at
\url{https://github.com/ph4r05/Whitebox-crypto-AES}.
The decryption primitive and its
embedded key are implemented by means of large tables that total
1.14MB. Renewing this routine and its tables to renew the decryption key hence
involves the downloading of a mobile block of about 1.14MB. This is
significantly larger than the code blocks that were downloaded in the bzip2
experiments.

\begin{table}[!t]
\caption{Client CPU consumption and wall-clock execution times of renewable WBC}
  \small
\centering
\begin{tabularx}{\textwidth}{c|YYY|YYY}
\textbf{refresh time (s)} & \multicolumn{3}{c|}{\textbf{user-space CPU time (s)}}       & \multicolumn{3}{c}{\textbf{wall-clock exec.\ time (s)}}                                                         \\ \hline
           & Mean & StDev & overhead  & Mean & StDev & overhead\\ \hline 

baseline    & 156.1 \ & 0.4 \ & -                  & 156.7 \ & 0.4         \        & -                     \\ \hline
1           & 161.0 \ & 0.5 \ & 3.1\%              & 179.0 \ & 0.6       \          & 14.2\%                      \\
2           & 158.8 \ & 0.4 \ & 1.7\%              & 165.6 \ & 0.4     \            & 5.6\%                     \\
3           & 157.4 \ & 0.6 \ & 0.8\%              & 162.5 \ & 0.6   \              & 3.7\%                      \\
4           & 156.9 \ & 0.5 \ & 0.5\%              & 160.6 \ & 0.5   \              & 2.5\%                         \\
5           & 156.3 \ & 0.6 \ & 0.1\%              & 159.8 \ & 0.5  \               & 2.0\%
\end{tabularx}
\label{table:wbc-user}
\end{table}

\begin{table}[!t]
\caption{Network throughput of renewable WBC}
\small
  \centering
\begin{tabularx}{0.8\textwidth}{c|YY|YY}
\textbf{refresh time (s)} & \multicolumn{2}{c|}{\textbf{transferred blocks}}                   & \multicolumn{2}{c}{\textbf{transferred MBs}}                            \\ \hline
         & {\textbf{Mean}} & {\textbf{StDev}} & {\textbf{Mean}} & {\textbf{StDev}} \\ \hline
1        & 179.8    & 0.7 \ \    & 205.1   & 0.8  \ \                                 \\
2        & 83.4     & 0.5 \ \    & 95.1    & 0.8  \ \                                 \\
3        & 54.9     & 0.4 \ \    & 62.6    & 0.4  \ \                                 \\
4        & 40.9     & 0.3 \ \    & 46.7    & 0.4  \ \                                 \\
5        & 32.5     & 0.5 \ \    & 37.0    & 0.6  \ \
\end{tabularx}
\label{table:wbc-network}
\end{table}

Table~\ref{table:wbc-user} reports client user-land CPU consumption times and client wall-clock execution
times of the baseline version without renewability, and of the renewable version
at different refresh rates. The
differences between the overheads in both measurements is \changed{considerable}. This is of
course due to the fact that the client side spends a significant amount of time
waiting for the large mobile blocks to arrive. However, during that wait, no CPU
resources are consumed. Still, even for the version that only refreshes
the routine and its embedded key every 5 seconds, the user-land CPU time
increases significantly. The reason is that the Downloader and Binder components
take up some computation time, and that the code transformations that are
necessary to implement code mobility and renewability---as detailed in our
previous work~\cite{cm}---also have a small, but significant effect on
performance.

Table~\ref{table:wbc-network} shows how the network throughput scales
with the refresh rates. The number of transferred blocks, which
equals the number of refreshes (plus 1) scales super-linearly with the refresh
frequency because the execution time of the benchmark increases with higher
refresh frequencies. For this form of renewability, which inherently involves
large mobile blocks, the measurements confirm that network scalability is an important issue to consider.

\iftrue
\section{Related Work}
\label{sec:related}

Our framework combines and extends concepts from network-based protections, and
software diversity. Network-based software protection techniques leverage
software updates and trusted network services. 
The updates may be implemented for the functional part of the program, and for
the protection techniques used to protect it~\cite{IEEESP13}. Both Collberg et
al.~\cite{collberg2012distributed} and Falcarin et
al.~\cite{falcarin2011exploiting} proposed the continuous replacement of binary
code. Collberg et al.\ make use of CIL (Common Intermediate Language) to
generate diversified code. They support both syntactic and semantic diversity,
using what they call Protocol-Preserving and Non-Protocol-Preserving
Transformation Primitives, respectively. \changed{They only diversify
application code in order to overwhelm an attacker with new code versions to
increase the required attacked effort. They consider no integration with other
protections or making other protections renewable. Furthermore, no code was made
available. Falcarin et al.\ only pitched the idea of making existing application
code mobile as a form of obfuscation and proposed binary rewriting as an
implementation option, but they provided no experimental validation or prototype
implementations, nor did they consider composability or renewability of other
protections.} Contrary to the work of Collberg et
al.~\cite{collberg2012distributed}, our framework works by directly replacing
binary code, giving it more freedom in terms of granularity and composability
with other techniques. Contrary to both, our framework not only makes it
possible to renew application code, it can also renew entire protection
techniques, in an automated, specialized, manner. Contrary to both, our
framework has also been validated by industrial experts~\cite{D106}.

Previous Java work implemented dynamic replacement of remote attestation
protection code downloaded by a trusted server, using extended Java Virtual
Machines~\cite{scandariato2008application}. Other techniques such as remote
attestation extend code guards with a network server.
The Pioneer~\cite{seshadri2007pioneer} system relied on a verification function running on the client as an OS primitive, and an attestation server. 
Garay et al.~\cite{garay2006software} presented an approach where a trusted challenger sends a challenge to the potentially corrupted responder. The challenge is an executable program that can execute any function on the responder, which must compute the challenge fast enough to prove its integrity.


In literature~\cite{franz2010unibus,jackson2011compiler,davi2012}, software diversity relied on random generation of diversified copies, starting from the same source code, extending the idea of compiler-guided code variance~\cite{forrest97}. 
A survey~\cite{Larsen2014} compares the different approaches for software diversity in terms of performance and security, and recently software diversity has become practical due to cloud computing enabling the computational power to perform massive diversification~\cite{Larsen2014}.
Past software diversity approaches have been based on some form of obfuscation~\cite{cohen1993operating}, load-time binary transformation~\cite{just2004review}, virtualization obfuscation based on customized virtual machines~\cite{holland2005architecture}, or OS randomization~\cite{xu2003transparent}.
Other approaches rely on binary transformation based on a random seed~\cite{diablo}, or multi-compilers and cloud computing~\cite{franz2010unibus} to create a unique diverse binary version of every program, and  they apply such diversification for mobile apps~\cite{jackson2011compiler}.
The XIFER framework~\cite{davi2012} randomly diversifies Android apps at load time by means of a binary rewriter. 
Both spatial and temporal software diversity has been proposed as a solution to
a wide range of problems: code randomization has been used to defend against
code-reuse attacks~\cite{Shioji2012}, return-oriented programming
attacks~\cite{gupta2013}, and code injection attacks~\cite{williams2009}. More
fine-grained forms of diversification have been proposed to raise the bar even
further~\cite{giuffrida2012,kil2006address}, including for code dynamically
generated with JIT compilers~\cite{homescu2013librando}. Dynamic temporal
diversity has been proposed to mitigate timing side channel attacks
~\cite{crane2015thwarting}. 
Diversification can also prevent collusion attacks to identify vulnerabilities~\cite{IEEESP13}.

\changed{With the work presented in this paper, we do not aim for pushing the state-of-the-art in terms of code diversification itself. We simply leverage the existing state-of-the-art in diversity techniques in support of renewability. For example, the syntactic diversification discussed in Section~\ref{app:syntax} reuses the diversification techniques already deployed to mitigate collusion attacks in~\cite{IEEESP13}. For WBC, e.g., any domain-specific technique to generated diversified instances can be used, as that choice is completely orthogonal to the rest of our framework.}


\changed{To the best of our knowledge, existing MATE software protection tools available to (academic) researchers, including Tigress, OLLVM, Sandmark, and ProGuard, support none of the forms of renewability we discussed in Section~\ref{sec:applications}. The three latter only support static obfuscation. While Tigress offers rather strong diversification in combination with static and dynamic obfuscations, it is limited to compile-time diversification: Once an obfuscated binary is distributed, it is fixed. The obfuscated binary may generate code itself by means of a just-in-time (JIT) compiler, and each distributed binary may generate different JIT-ed code, but the code JIT-ed by a specific distributed binary is never renewed or diversified.}

Compared to the discussed work, our renewability framework
provides a foundation to combine, compose, extend, and hence fortify several
existing defenses (beyond mere obfuscation). The tool flow supports combinations and compositions, meaning
that multiple protections can be deployed together on the same program or even
on the same code fragment. This follows in part from its conception as part of
the ASPIRE Compiler Tool Chain, the software protection tool chain developed in
the ASPIRE project as automated support for a wide range of software
protections. Our framework is fully
compliant with the ASPIRE software protection reference
architecture~\cite{wicsa2016reference,D1.04}. 
 
As demonstrated, our framework is applicable to native code, and is hence not limited to
code distributed in higher-level, more symbolic (and hence easier to attack)
formats such as Java bytecode. The granularity of the renewability is
furthermore not limited to coarse code fragments such as whole functions. Much
smaller (security-sensitive) code regions can instead be made renewable.

As already discussed in Sections~\ref{sec:attack_model}
and~\ref{sec:applications}, the framework and concrete instantiations of its
capabilities can mitigate concrete attack paths.
Recently, Ceccato et al.\ reported results of a qualitative analysis of how
professional hackers as well as amateurs understand protected code while
performing attack steps~\cite{emse2019}. The resulting taxonomy of concepts used
by the hackers to describe their attacks towards code understanding, and the
inferred models of their activities and their reasoning, provide further
insights into how the proposed renewability framework can impede certain attack
paths and attack strategies. Several activities are impacted by renewability as
supported by our architecture and tool flow, including but not limited to:
static analysis, tracing, debugging, statistical analysis, assessing the effort,
building of workarounds, undoing of protections, overcoming of protections,
formulating hypotheses, and confirmation of hypotheses. The latter two play an
important role in real-world attacks. They depend to a large degree on
repeatability of attack activities, which is directly addresses by the forms of
renewability our framework supports.

\fi

\changed{There are plenty of commercial protection tools available on the market,
such as those from Arxan, Irdeto, and Guardsquare. Those are notably
missing in the above discussion. The reason is that the commercial companies are
very secretive. For example, the academics authors of this paper cannot get
access to accurate enough documentation (i.e., beyond the level of marketing
info) to allow scientifically solid comparisons.  What we do know is that
some commercial protection tool suites deploy source-level and binary-level
protections. Furthermore, their deployment of all kinds of obfuscations is
syntactically diversified, in the sense that the static code generated for a
protection looks different every time to mitigate the simplest attack vectors
such as pattern matching. In that regard our tool suite is not novel. We don't
feel confident, however, writing about more specific code mobility or more
dynamic renewability capabilities or features of the commercial offerings.

Whether their deployment is supported by tools or done manually in an ad-hoc
manner is unclear because of the already mentioned secrecy, but do we know of at
least three forms of renewability that are used commercially. The first one is
the evolution of license key checks as discussed in
Section~\ref{app:evolving}. To the best of our knowledge, that is deployed only
in a static manner, i.e., with extended checks embedded statically in
consecutive software released. The second form is an ad-hoc instantiation of the
aforementioned idea by Collberg et al.\ to overwhelm attackers with new code
version~\cite{collberg2012distributed}. In some live video distribution schemes
(e.g., via satellite pay TV) decompression and decryption code is sent along with
the video streams and is renewed frequently, i.e., up to multiple times per
second. As the value of live content drops really quickly, such a protection
implies that in order to be successful, attackers must crack each version within
seconds to minutes, and they must do so for tens to hundreds of versions in
parallel. In practice, this proves to provide strong protection. This form of
protection is supported by our framework as discussed in Section~\ref{app:wbc},
albeit that the renewed code is not embedded in the streamed data in our
framework. Which form provides the best protection is an open research question
at the moment. A third commercially used form is VM-based renewability in which,
e.g., Lua scripts that check the integrity of the installation on a player's
computer to prevent cheating are renewed on a regular basis in online games.}

\section{Availability}
\changed{The ASPIRE Compiler Tool Chain is available as open-source at and via
  \url{https://github.com/aspire-fp7/framework}. This includes the link-time
  rewriter, the compiler patches, the mobility and renewability server and
  client components, scripts to invoke the whole tool flow and to handle
  source-code annotations, and extensive documentation that is available at
  \url{https://aspire-fp7.eu/}. The open sourced code includes many concrete
  protections, such as remote attestation, anti-debugging by means of
  self-debugging, control flow obfuscation, code guards, anti-callback stack
  checks, etc. It excludes some protections that were researched in the ASPIRE
  project but that were only developed in proprietary plug-in prototypes, such
  as the white-box cryptography that was deployed in the industrial Android use cases, data obfuscation, and instruction set virtualization. The
  Android use cases are not available either, due to their inclusion of
  security-sensitive assets from the partnering companies. The bzip2 benchmark and the smaller white-box cryptography benchmark we deployed are available however. About 4 hours of
  video demonstrations of the whole ASPIRE tool chain, including the renewability
  framework, have been published in the ASPIRE project YouTube channel at
  \url{https://goo.gl/pfESbK}.}

\changed{By making this tool chain available, we aim to provide useful
  infrastructure for future research. In the domain of software protection,
  there is a large discrepancy between what companies do and have available in
  secret, and what academics have available to experiment with in public. The
  lack of defenders to evaluate how new contributions compose with the existing
  state-of-the-art hampers progress. By providing research infrastructure, we
  hope the software protection community can catch up with, e.g., the domain of
  cryptography, where there is a constant back and forth of attacks and
  defenses. This is absolutely necessary, as software protection is bound to
  remain part of the never-ending arms race between defenders and attackers.}

\section{Conclusions}
\label{sec:conclusions}
This paper presented the ASPIRE framework, architecture and tool flow support
for native code renewability. This framework supports several forms of
renewability, in which renewed and diversified code and data, belonging to either
the original application or to linked-in protection components, is delivered from
a secure server to a client application on demand. This results in frequent
changes to the software components when they are under attack, thus making
dynamic attacks harder. Several applications of the renewability framework have
been discussed, some of which extend existing protections, and some of which
enforce existing protections. The prototype implementation was evaluated
successfully on a number of use cases, including complex libraries
representative for real-world, industrial use cases. 
Most of the prototype implementations are available online as open source.


\bibliographystyle{ACM-Reference-Format}
\bibliography{references}

\end{document}